\documentclass[12pt]{article}
\usepackage{graphicx}

\def\ignore#1{{}}

\let\oldtheequation=\theequation
\def\doteqs#1{\setcounter{equation}{0}            
\def\theequation{{#1}.\oldtheequation}}
\newcounter{sxn}
\def\sx#1{\addtocounter{sxn}{1} \vskip 1.cm  \goodbreak
\noindent{\large\bf\leftline{\thesxn.~~#1}} \nobreak \vskip -.5cm}
\def\sxn#1{\sx{#1} \doteqs{\thesxn}}

\newcounter{axn}

\date{}

\newdimen\mybaselineskip
\newcommand{\beeq}{\begin{equation}}
\newcommand{\eneq}{\end{equation}}
\newcommand{\beqn}{\begin{eqnarray}}
\newcommand{\eeqn}{\end{eqnarray}}

\def\la{\raise.16ex\hbox{$\langle$}\lower.16ex\hbox{}  }
\def\ra{\, \raise.16ex\hbox{$\rangle$}\lower.16ex\hbox{} }

\def\psibar{ \psi \kern-.65em\raise.6em\hbox{$-$} \lower.6em\hbox{} }
\def\psibarb{ \psi \kern-.65em\raise.6em\hbox{$-$}  }

\begin{document}

\thispagestyle{empty}

\baselineskip=12pt



\vspace*{3.cm}

\begin{center}  
{\LARGE \bf  The Highly Damped Quasinormal Modes of Extremal Reissner-Nordstr$\ddot{\rm o}$m  and Reissner-Nordstr$\ddot{\rm o}$m-de Sitter Black Holes}
\end{center}

\baselineskip=14pt

\vspace{3cm}
\begin{center}
{\bf  Ramin G. Daghigh$\sharp$ and Michael Green$\dagger$}
\end{center}

\centerline{\small \it $\sharp$ Natural Sciences Department, Metropolitan State University, Saint Paul, Minnesota, USA 55106}
\vskip 0 cm
\centerline{} 

\centerline{\small \it $\dagger$ Applied Mathematics Department, Metropolitan State University, Saint Paul, Minnesota, USA 55106}
\vskip 0 cm
\centerline{} 

\vspace{1cm}
\begin{abstract}
We analyze in detail the highly damped quasinormal modes of $D$-dimensional extremal Reissner-Nordstr$\ddot{\rm{o}}$m and Reissner-Nordstr$\ddot{\rm{o}}$m-de Sitter black holes.  We only consider the extremal case where the event horizon and the Cauchy inner horizon coincide. 
We show that, even though the topology of the Stokes/anti-Stokes lines in the extremal case is different than the non-extremal case, the highly damped quasinormal mode frequencies of extremal black holes match exactly with the extremal limit of the non-extremal black hole quasinormal mode frequencies. 

\baselineskip=20pt plus 1pt minus 1pt
\end{abstract}

\newpage

\sxn{Introduction}

\vskip 0.3cm

It is believed that black holes will play a crucial role in the discovery of the laws of a quantum theory of gravity just as the hydrogen atom did in the evolution of quantum mechanics \cite{Bekenstein}.  This issue motivates us to study the quasinormal mode (QNM) spectra of black holes.  These modes are classical waves emitted by a perturbed black hole.
In recent years, the highly damped QNMs of black holes have attracted a great deal of attention due to Hod's proposal\cite{Hod} that these modes may be providing information about quantum gravity and in particular the quantization of the horizon area of black holes.  This proposal was in part motivated by the fact that the real part of the QNM frequency ($\omega_R$) approaches a universal value for Schwarzschild black holes in the highly damped limit with the special numerical value of
\beeq
\omega_R \mathop{\longrightarrow}_{|\omega_I|\to\infty}\ln(3)kT_{bh}/\hbar ~,
\label{ln3}
\eneq
where $\omega_I$ is the damping and $T_{bh}$ is the Hawking temperature of the black hole.  The special numerical value $\ln(3)$ allows an elegant statistical interpretation\cite{Hod, Bekenstein-1} of the resulting Bekenstein-Hawking entropy spectrum. This special value of $\ln(3)$ appears for a large class of single horizon black holes\cite{Motl2,Gabor3,Das1,Ramin1}.  The simple relationship (\ref{ln3}) breaks down in multi-horizon black holes\cite{mh-1,mh-2,mh-3,Andersson,mh-4,Natario-S,Daghigh-RN}.  For a possible explanation on why the relationship (\ref{ln3}) does not hold for Reissner-Nordstr$\ddot{\rm{o}}$m (RN) black holes, one can refer to \cite{Andersson}.

In this paper we focus on the highly damped QNMs of extremal RN black holes.  These black holes hold an important and controversial status in black hole physics.  Traditionally these black holes were believed to be a limiting case of non-extremal black holes\cite{Misner}.  This traditional view was challenged in \cite{Hawking-Horowitz} based on the fact that the topology of the extremal and non-extremal black holes have qualitative differences.  Based on these differences, the authors of \cite{Hawking-Horowitz} and \cite{Teitelboim} argued that extremal RN black holes have zero entropy with no definite temperature despite having a non-zero horizon area.  These black holes are also important in the context of supergravity theories\cite{Supergravity}.  It has been shown in \cite{Horowitz} that an exact solution of these black holes exists in (super)string theory.  

Regarding the QNM spectra of extremal RN black holes, not much can be found in the literature.  In \cite{Onozawa1}, the authors have computed the least-damped modes (contrary to the case considered in this paper) for four dimensional extremal RN black holes.  These authors found that the QNM frequencies of gravitational waves with multi-pole index $l$ and electromagnetic waves with multi-pole index $l-1$ coincide.  Based on this observation, the authors of \cite{Onozawa2}  conjectured that the modes of different perturbations can be matched because of the supersymmetry transformations in the extremal solution.  The numerical calculation of the QNM frequencies of extremal RN black holes in four spacetime dimensions has been done by Berti in \cite{Berti1}.  Berti's results show that the extremal RN QNM frequencies have a similar pattern to Schwarzschild frequencies.  As the damping increases the frequency seems to approach the same constant value as in Schwarzschild black holes where
\beeq
\omega_R \mathop{\longrightarrow}_{|\omega_I|\to\infty}\ln(3)c^3/8\pi G M ~.
\label{ln3-extreme}
\eneq
This result seems to be compatible with the extremal limit of the non-extremal RN black hole QNM frequency which was derived in \cite{Andersson}. Unfortunately the numerical method in \cite{Berti1} is not stable in the highly damped limit in order to explicitly
verify the result in (\ref{ln3-extreme}).  It is suggested in \cite{Andersson} that since the topology of the Stokes/anti-Stokes lines for extremal RN black holes is different than the non-extremal case, the QNMs for these black holes would require a separate analysis.  Following this suggestion, the authors of \cite{Das1} and \cite{Natario-S} have attempted to explicitly calculate the highly damped QNM frequency of extremal RN black holes using the monodromy method of \cite{Motl2}.  Unfortunately, the results in both papers are wrong.  In \cite{Das1}, the authors have used an incorrect topology of the anti-Stokes lines.  Specifically, in the contour followed by the authors along the anti-Stokes lines, the rotation angle taken near the origin of the complex plane to move from one anti-Stokes line, which extends to infinity in the complex plane, to the other anti-Stokes line, which also extends to infinity in the complex plane, is incorrect.  For example, in \cite{Das1} the rotation angle in four spacetime dimensions is taken to be $\pi$ in the complex $r$-plane while the correct angle should be $5\pi/3$.  The authors of \cite{Natario-S} have used the correct topology, but in their calculations they have crossed the anti-Stokes lines which connect to the horizon.  One should remember that the WKB solutions are fixed on these lines due to the boundary condition at the horizon and crossing these lines without applying the boundary condition will lead to incorrect results.

The paper is organized as follows.  In Section 2, we describe the general formalism.  In Section 3, we calculate the highly damped QNM frequencies of extremal RN black holes in $D\ge 4$ spacetime dimensions using the analytic method of Andersson and Howls\cite{Andersson}.  In section 4, we extend our QNM calculations to include Reissner-Nordstr$\ddot{\rm{o}}$m-de Sitter (RN-dS) black holes in $D\ge 4$ spacetime dimensions.  Finally, in Section 5 we end the paper with some conclusions and remarks.

\sxn{General Formalism and the Highly Damped Limit}
\vskip 0.3cm

Ishibashi and Kodama have shown in \cite{Ishibashi1}, \cite{Ishibashi2}, and \cite{Ishibashi3} that various classes of non-rotating black hole metric perturbations in a spacetime with dimension $D>3$ are governed generically by a Schr$\ddot{\mbox o}$dinger wave-like equation of the form
\beeq
{d^2\psi \over dz^2}+\left[ \omega^2-V(r) \right]\psi =0 ~,
\label{Schrodinger}
\eneq
where $V(r)$ is the QNM potential obtained by Ishibashi and Kodama.  Here, the perturbations depend on time as $e^{-i\omega t}$.  The Tortoise coordinate $z$ is defined by 
\beeq
dz ={dr \over f(r) }~,
\label{tortoise}
\eneq
where $f(r)$ is related to the spacetime geometry, and is given by 
\beeq
f(r)=1-{2\mu \over r^{D-3}}+{\theta^2 \over r^{2D-6}}-\lambda r^2~.
\label{function f}
\eneq
The ADM mass, $M$, of the black hole is related to the parameter $\mu$ by
\beeq
M = {(D-2)A_{D-2} \over 8 \pi G_D}\mu~,
\eneq
where $A_n$ is the area of a unit $n$-sphere,
\beeq
A_n={2\pi^{n+1 \over 2} \over \Gamma\left({n+1 \over 2}\right) }~.
\eneq
The electric charge, $q$, of the black hole is
\beeq
q^2 = { (D-2)(D-3)\over 8 \pi G_D }\theta^2~,
\eneq
while the value of the cosmological constant, $\Lambda$, is given by 
\beeq
\Lambda = { (D-1)(D-2)\over 2 }\lambda~.
\eneq

The effective potential $V(r)$, was found explicitly by Ishibashi and Kodama\cite{Ishibashi1,Ishibashi2,Ishibashi3} for scalar (reducing to polar at $D=4$), vector (reducing to axial at $D=4$), and tensor (non-existing at $D=4$) perturbations.  The effective potential for tensor perturbations has been shown \cite{Gibbons, Konoplya} to be equivalent to that of the decay of a test scalar field in a black hole background. The effective potential is zero at both the event horizon ($z\rightarrow -\infty$) and spatial infinity (or cosmological horizon for asymptotically de Sitter space) ($z\rightarrow \infty$).  In the case of QNMs, the asymptotic behavior of the solutions is chosen to be
\beeq
\psi(z) \approx \left\{ \begin{array}{ll}
                   e^{-i\omega z}  & \mbox{as $z \rightarrow -\infty$ }~,\\
                   e^{+i\omega z}  & \mbox{as $z\rightarrow \infty$ ~,}
                   \end{array}
           \right.        
\label{asymptotic}
\eneq 
which represents an out-going wave at infinity (or cosmological horizon for asymptotically de Sitter space)and an ingoing wave at the event horizon. 

Since the tortoise coordinate is multi-valued, it is more convenient to work in the complex $r$-plane.  After rescaling the wavefunction $\psi=\Psi/\sqrt{f}$ we obtain
\beeq
\frac{d^2\Psi}{dr^2}+R(r)\Psi=0 ~,
\label{Schrodinger-r}
\eneq 
where
\beeq
R(r)= {\omega^2\over f^2(r)}-U(r)~,
\label{Rr}
\eneq
with
\beeq
U(r)={V(r)\over f^2}+{1\over 2}\frac{f''}{f}-\frac{1}{4}\left(\frac{f'}{f}\right)^2 ~.
\label{Ur}
\eneq
Here prime denotes differentiation with respect to $r$.


In the highly damped limit, where
\beeq
\omega \sim i~({\rm Im}~\omega) \to -i\infty~,
\eneq
it has been shown in \cite{Daghigh-RN} that when the cosmological constant is zero the potential $U(r)$ is negligible everywhere on the complex plane except when $r\rightarrow 0$.  In this region
\beeq
U(r)\sim \frac{J^2-1}{4r^2}~,
\label{Rr2}
\eneq
where $J$ is given by
\beeq
J= \left\{ \begin{array}{ll}
                   0~~~ \mbox{Tensor Perturbation ($\theta=0, n>2$)}~,\\
                   n-1~~~ \mbox{Tensor Perturbation ($\theta \neq 0, n>2$)}~,\\
                   2n~~~ \mbox{Vector Perturbation ($\theta = 0$)}~,\\
                   3n-1~~~ \mbox{Vector Perturbation ($\theta\neq 0$)}~,\\
                   0~~~ \mbox{Scalar Perturbation ($\theta= 0$)}~,\\
                   n-1~~~ \mbox{Scalar Perturbation ($\theta \neq 0$)}~,\\
                   \end{array}
           \right.        
\label{J}
\eneq
with $n=D-2$.  In the presence of a cosmological constant, the potential $U(r)$ also becomes important in the region where $r \rightarrow \infty$.  In this region 
\beeq
U(r)\sim {V(r) \over f^2} \sim \frac{J_{\infty}^2-1}{4r^2}~,
\label{Rr3}
\eneq
where $J_{\infty}$ is given by
\beeq
J_{\infty}= \left\{ \begin{array}{ll}
                   n+1~~~ \mbox{Tensor Perturbation}~,\\
                   
                   n-1~~~ \mbox{Vector Perturbation}~,\\
                 
                   n-3~~~ \mbox{Scalar Perturbation}~.\\
                   \end{array}
           \right.        
\label{Jinfinite}
\eneq
Equations (\ref{Rr3}) and (\ref{Jinfinite}) are valid for both charged ($\theta\ne 0$) and uncharged ($\theta = 0$) spherically symmetric black holes.

\sxn{Extremal Reissner-Nordstr$\ddot{\rm o}$m Quasinormal Modes}
\vskip 0.3cm

In order to find the highly damped QNM frequency of extremal RN black holes, we use the WKB approximation where the two solutions to Eq. (\ref{Schrodinger-r}) are given by
\beeq
\left\{ \begin{array}{ll}
                   f_1^{(t)}(r)=\frac{1}{\sqrt {Q(r)}}e^{+i\int_{t}^rQ(r') \, dr'}~,\\
                   \\
                   f_2^{(t)}(r)={1 \over \sqrt{Q(r)}}e^{-i\int_{t}^rQ(r')dr'}~.
                   \end{array}
           \right.        
\label{WKB}
\eneq
Here
\beeq
Q(r)=\sqrt{R(r)-\frac{1}{4r^2}}\sim \sqrt{\frac{\omega^2}{f^2}-\frac{J^2}{4r^2}}~
\label{Q^2-general}
\eneq
is shifted by $1/(4r^2)$ in order to guarantee the correct behavior of the WKB solutions at the origin\cite{Andersson}.\footnote{In the presence of a cosmological constant, it turns out that the shift of $1/(4r^2)$ is also necessary to guarantee the correct behavior of the WKB solutions at infinity.  The proof is almost identical to the proof shown in \cite{Andersson} for the region near $r = 0$ and therefore is not presented in this paper again.}   In Eq. (\ref{WKB}), $t$ is a simple zero of the function $Q^2$.  It is important to point out that the function $Q$ is multi-valued because of the square-root.  To make $Q$ single-valued, we have to introduce branch cuts from the simple zeros of $Q^2$.  As in \cite{Andersson}, we choose the phase of $Q$ such that the out-going wave solution at infinity (or cosmological horizon for asymptotically de Sitter space) is proportional to $f_1$ while the in-going wave solution at the horizon is proportional to $f_2$.

\begin{figure}[tb]
\begin{center}
\includegraphics[height=8cm]{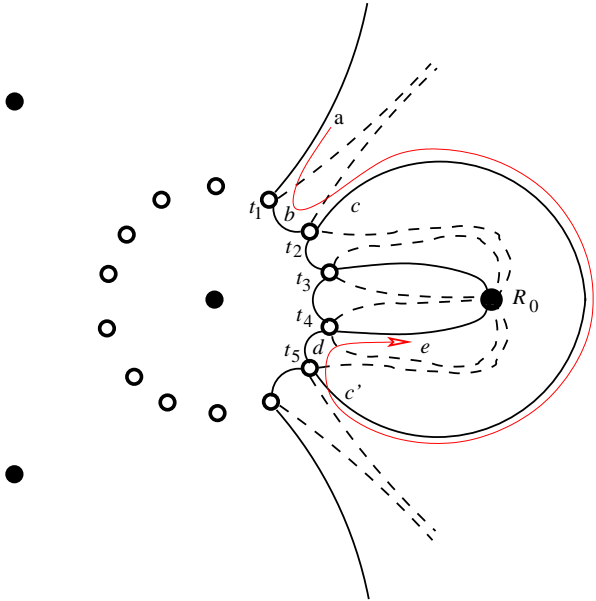}
\end{center}
\caption{Schematic presentation of the Stokes (dashed) and anti-Stokes (solid) lines in the complex plane for the RN solution in six spacetime dimensions.  The open circles are the zeros and the filled circles are the poles of the function $Q^2$.  The thin arrow which starts on the anti-Stokes line labeled $a$ and ends on the line that is connected to the extremal horizon (line $e$) is the path we take to determine the QNM frequency.}
\label{schem-d6}
\end{figure}

Given the function $Q$, we can determine the WKB condition for the highly damped QNM frequency. The methodology that we adopt is explained in \cite{Andersson}. First, we determine the zeros and poles of the function $Q$ and consequently the behavior of the Stokes and anti-Stokes lines in the complex $r$-plane.  Stokes lines are the lines on which the WKB phase ($\int Q dr$) is purely imaginary and anti-Stokes lines are the lines on which the WKB phase is purely real.  In Fig. \ref{schem-d6} we show schematically a portion of the topology of the Stokes and anti-Stokes lines, which is relevant to our calculations, for the extremal RN solution in six spacetime dimension.  To see the topology of the anti-Stokes lines for these black holes in other spacetime dimensions, the reader can refer to \cite{Natario-S} where it is shown that in any spacetime dimension the portion of the topology shown in Fig. \ref{schem-d6} is generic in the sense that there are always two unbounded anti-Stokes lines which extend to infinity on either side of a bounded anti-Stokes line that encircles the horizon, and inside the bounded anti-Stokes line there are two anti-stokes lines which end at the horizon.  After determining the topology of the Stokes/anti-Stokes lines, we follow the path shown in Fig. \ref{schem-d6} starting on an unbounded anti-Stokes line on line $a$ and ending on an anti-Stokes line which connects to the horizon (line $e$).  Note that, since the topology shown in Fig. \ref{schem-d6} is generic, the following results hold for arbitrary spacetime dimensions greater than three.  The solution on line $a$ is known due to the boundary condition at infinity:
\beeq
\Psi_{a}=f_1^{(t_1)}~.
\eneq 
We apply the rules that are explained by Andersson and Howls\cite{Andersson} in moving along the anti-Stokes lines.  The steps we take are as follows:
\beeq
\Psi_b = f_1^{(t_1)} - if_2^{(t_1)} =e^{i\gamma_{12}} f_1^{(t_2)} - ie^{-i\gamma_{12}} f_2^{(t_2)}
\eneq
\beeq
\Psi_c = e^{i\gamma_{12}} f_1^{(t_2)} - i \left(e^{i\gamma_{12}} + e^{-i\gamma_{12}} \right) f_2^{(t_2)}
\eneq
\beeq
\Psi_{c'} = e^{i\gamma_{12}}e^{i\tilde{\gamma}_{25}} f_1^{(t_5)} - i \left(e^{i\gamma_{12}} + e^{-i\gamma_{12}} \right) e^{-i\tilde{\gamma}_{25}} f_2^{(t_5)}
\eneq
\begin{eqnarray}
\Psi_{d} &=& \left[ e^{i\gamma_{12}}e^{i\tilde{\gamma}_{25}} + \left(e^{i\gamma_{12}} + e^{-i\gamma_{12}} \right) e^{-i\tilde{\gamma}_{25}} \right] f_1^{(t_5)} - i \left(e^{i\gamma_{12}} + e^{-i\gamma_{12}} \right) e^{-i\tilde{\gamma}_{25}} f_2^{(t_5)}\nonumber \\
&=& \left[ e^{i\gamma_{12}}e^{i\tilde{\gamma}_{25}} + \left(e^{i\gamma_{12}} + e^{-i\gamma_{12}} \right) e^{-i\tilde{\gamma}_{25}} \right] e^{i\gamma_{54}}f_1^{(t_4)} \nonumber \\
&& - i \left(e^{i\gamma_{12}} + e^{-i\gamma_{12}} \right) e^{-i\tilde{\gamma}_{25}} e^{-i\gamma_{54}}f_2^{(t_4)}
\end{eqnarray}
\begin{eqnarray}
\Psi_{e} & = & \left[ e^{i\gamma_{12}}e^{i\tilde{\gamma}_{25}} e^{i\gamma_{54}} + \left(e^{i\gamma_{12}} + e^{-i\gamma_{12}} \right) e^{-i\tilde{\gamma}_{25}} e^{i\gamma_{54}} + \left(e^{i\gamma_{12}} + e^{-i\gamma_{12}} \right) e^{-i\tilde{\gamma}_{25}} e^{-i\gamma_{54}} \right] f_1^{(t_4)} \nonumber \\
 & &  
  - i \left(e^{i\gamma_{12}} + e^{-i\gamma_{12}}\right) e^{-i\gamma_{25}} e^{-i\gamma_{54}} f_2^{(t_4)}~
\end{eqnarray}
where
\beeq
\gamma_{ij} =\int_{t_i}^{t_j} Q dy \approx  \int_{t_i}^{t_j} \left[\frac{r^{4(n-1)}}{\theta^4}\omega^2-\frac{J^2}{4r^{2}}\right]^{1/2}dr ~
\label{gamma1}
\eneq
and
\beeq
\tilde{\gamma}_{25} = \Gamma - \gamma_{54} - \gamma_{43} - \gamma_{32}~.
\label{gamma25}
\eneq
Here $t_i$ and $t_j$ are the simple zeros of the function $Q^2$, $\tilde{\gamma}_{25}$ is the integral of $Q$ along the anti-Stokes line to the right of the horizon at $R_0$, and $\Gamma$ is the integral of $Q$ along a closed loop around the horizon in the clockwise direction.
We can solve the integral (\ref{gamma1})  by introducing a new variable $\eta=2\omega r^{2n-1}/J\theta^2$ which maps the zeros to $-1$ or $+1$, and this integral becomes
\beeq
\gamma_{ij}= \frac{J}{2(2n-1)} \int_{\mp1}^{\pm 1} \left(1-\frac{1}{\eta^2}\right)^{1/2}d\eta=\mp \frac{J}{2(2n-1)}\pi~.
\label{gamma2}
\eneq
It is now easy to show that
\beeq
\gamma = -\gamma_{12} = -\gamma_{32} = \gamma_{43} = -\gamma_{54} = \frac{J}{2(2n-1)}\pi~.
\label{gamma12}
\eneq
Since the line labeled $e$ is connected to the horizon, on this line we need to impose the boundary condition where we have purely in-going waves.  (This is where the problem occurs in \cite{Natario-S}.  There, the authors apply the boundary condition to the coefficients only after returning to the starting point and having twice crossed anti-Stokes lines that connect to the horizon.  However, the boundary condition needs to be applied as soon as one reaches an anti-Stokes line connected to the horizon.)  This means that the coefficient of $f_1$, which represents an out-going wave in our choice of phase for $Q$, must be zero.  By setting the coefficient of $f_1$ to zero and using Eqs. (\ref{gamma25}) and (\ref{gamma12}), we arrive at the WKB condition on highly damped QNMs:
\beeq
e^{2i\Gamma} = -2 - 2\cos(2\gamma) = -2 - 2\cos\left(\frac{J}{2n-1}\pi\right)~.
\label{RN-WKB}
\eneq
The constant $J$ for tensor and scalar perturbations is $n-1$, while for vector perturbations it is $3n - 1$ according to Eq. (\ref{J}).  However, even though  $J$ is different in vector perturbations, we can write
\beeq
{J_{vector} \over 2n-1}={3n-1 \over 2n-1}=2-{n-1 \over 2n-1}~,
\eneq 
which produces the same result once it is plugged into Eq. \ref{RN-WKB}. 
In other words, for all types of metric perturbations one can write 
\beeq
\cos\left({J \over 2n-1}\pi\right)= \cos\left({n-1 \over 2n-1}\pi\right)~.
\label{J-gen}
\eneq 
Using the residue theorem we evaluate $\Gamma$
\beeq
\Gamma=\oint Q dr= -2\pi i \mathop{Res}_{r=R_0} Q= -2\pi i\left[{\frac{n}{(n-1)^2}\mu^{1/(n-1)}\omega}\right] ~,
\label{Gamma}
\eneq
and finally we can determine the highly damped QNM frequency $\omega$ to be 
\beeq
{\frac{4\pi n}{(n-1)^2}\mu^{1\over {n-1}}\omega} = \ln\left[2 + 2\cos\left(\frac{n-1}{2n-1}\pi\right)\right]+(2k+1)\pi i~~\mbox{as $k\rightarrow \infty$~,}
\label{omega_final}
\eneq
where $k$ is an integer.  This result is valid for all types of metric perturbations in four and higher dimensions.  Note that $\ln(3)$ only appears in four spacetime dimensions.  In higher spacetime dimensions, we get the the natural logarithm of a real number between $2$ and $3$.

Clearly this result agrees with the extremal limit of the highly damped QNM frequencies of non-extremal RN black holes which was derived by Andersson and Howls\cite{Andersson} for four spacetime dimensions and by Natario and Schiappa\cite{Natario-S} for all spacetime dimensions greater than three.

\begin{figure}[tb]
\begin{center}
\includegraphics[height=3.5cm]{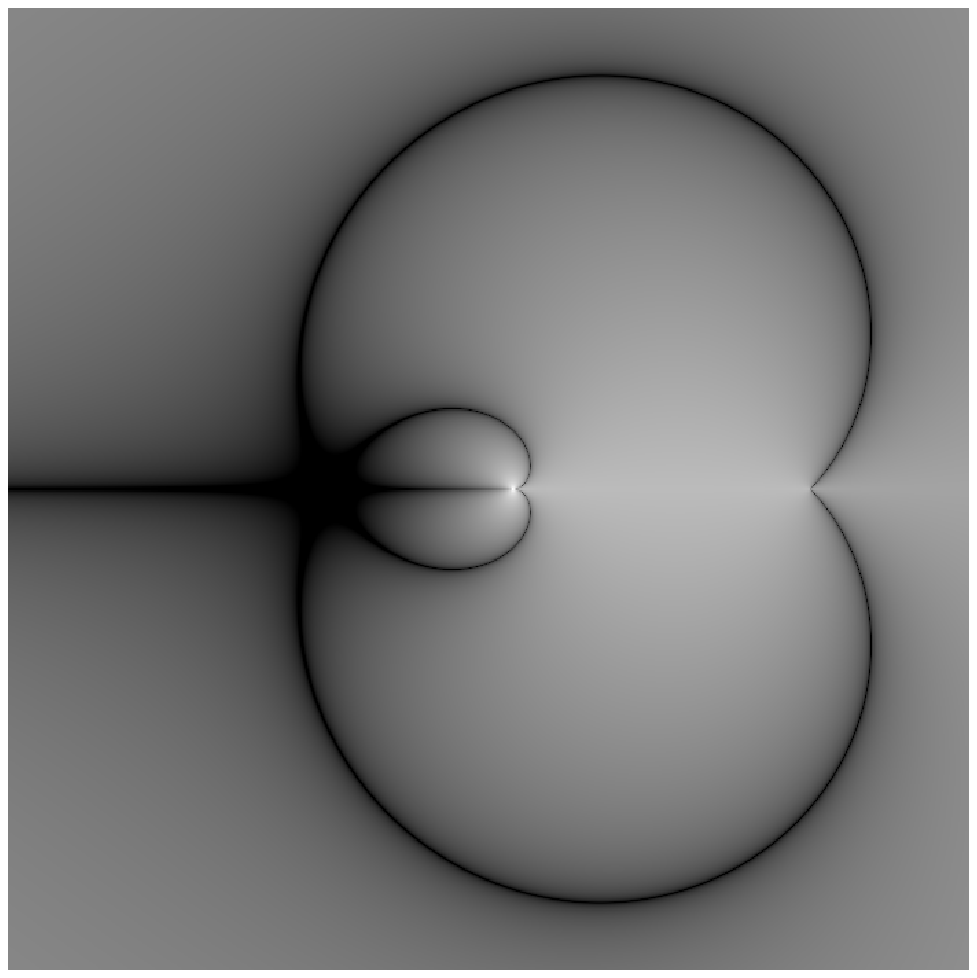}
\includegraphics[height=3.5cm]{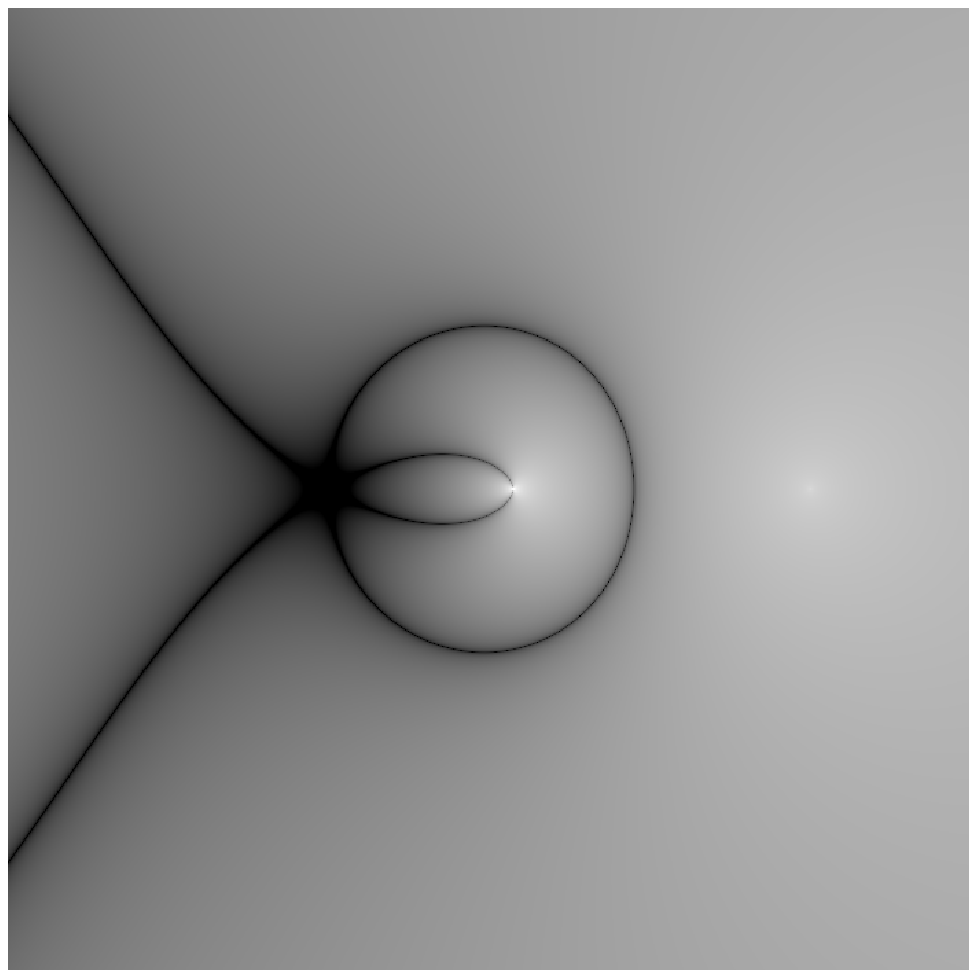}
\includegraphics[height=3.5cm]{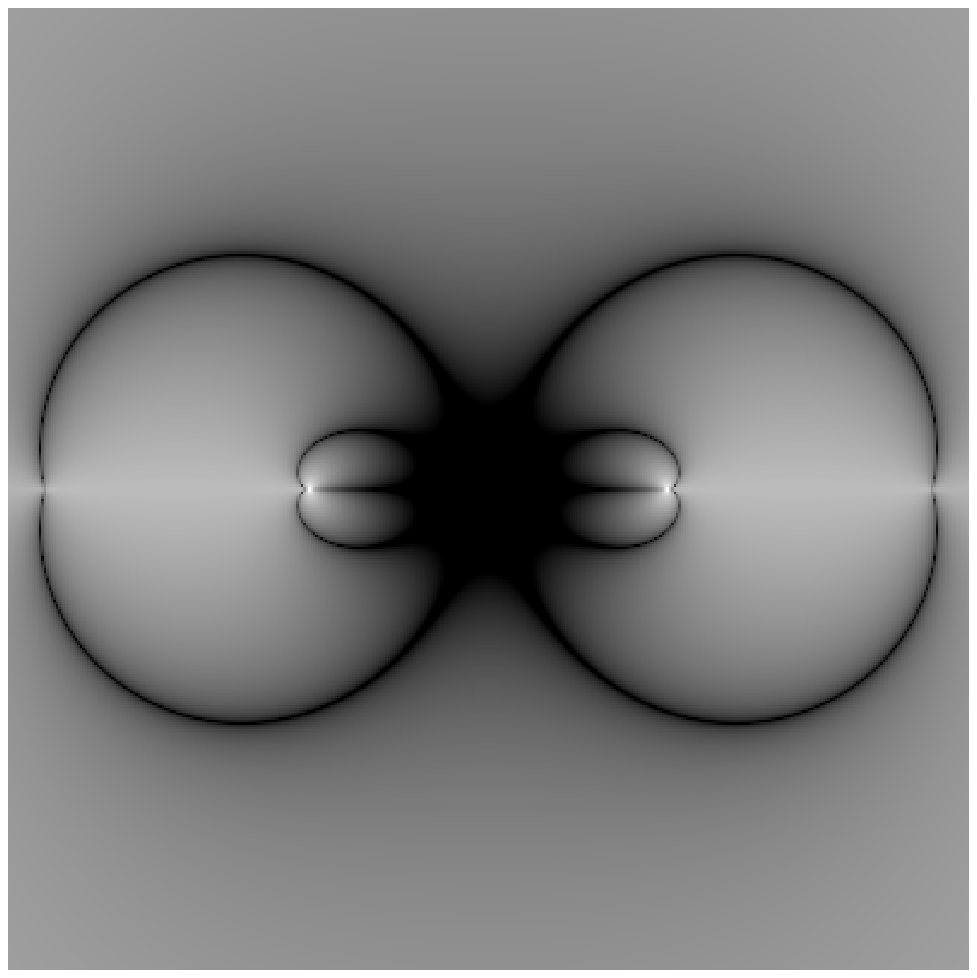}
\includegraphics[height=3.5cm]{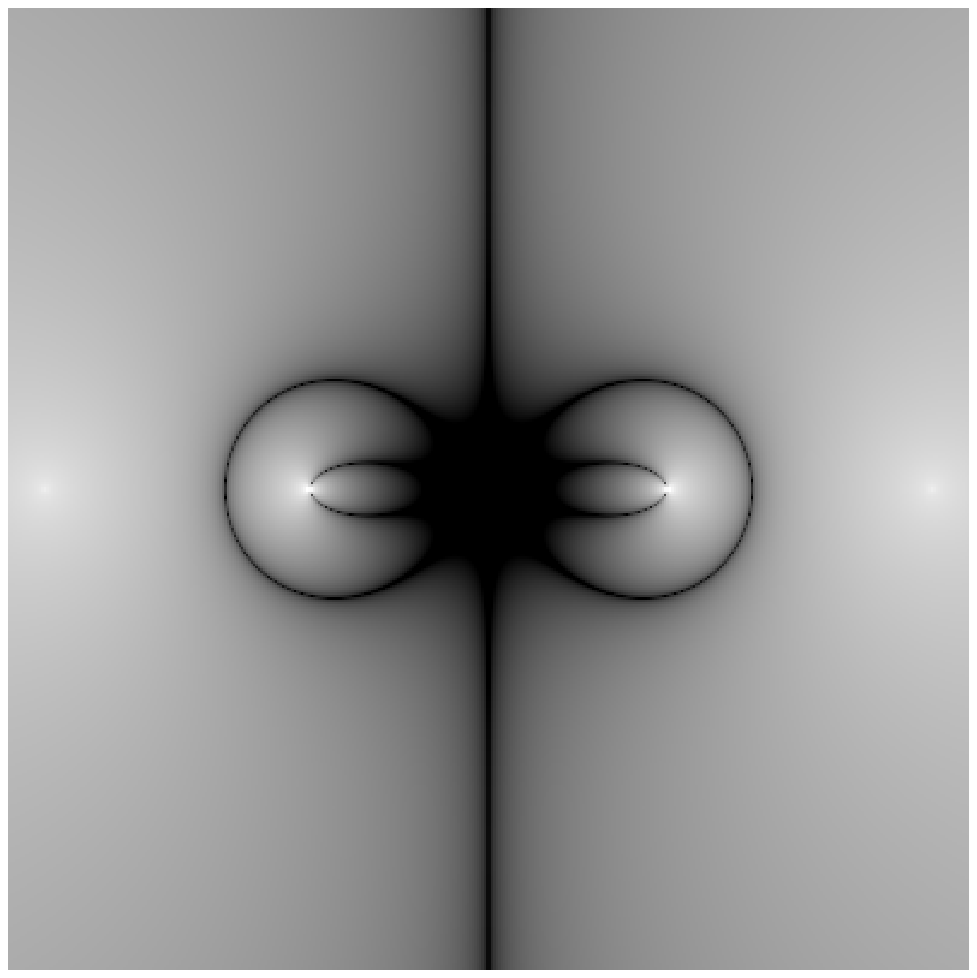}
\end{center}
\caption{Numerical calculation of the Stokes and anti-Stokes lines for extremal RN-dS black holes.  From left to right we show the Stokes and anti-Stokes lines in $D=4$, then the Stokes and anti-Stokes lines in $D=5$.  For the complete structure, see Fig. \ref{schem-d4-dS} and Fig. \ref{schem-d5-dS}.}
\label{numerical}
\end{figure}

\begin{figure}[tb]
\begin{center}
\includegraphics[height=7cm]{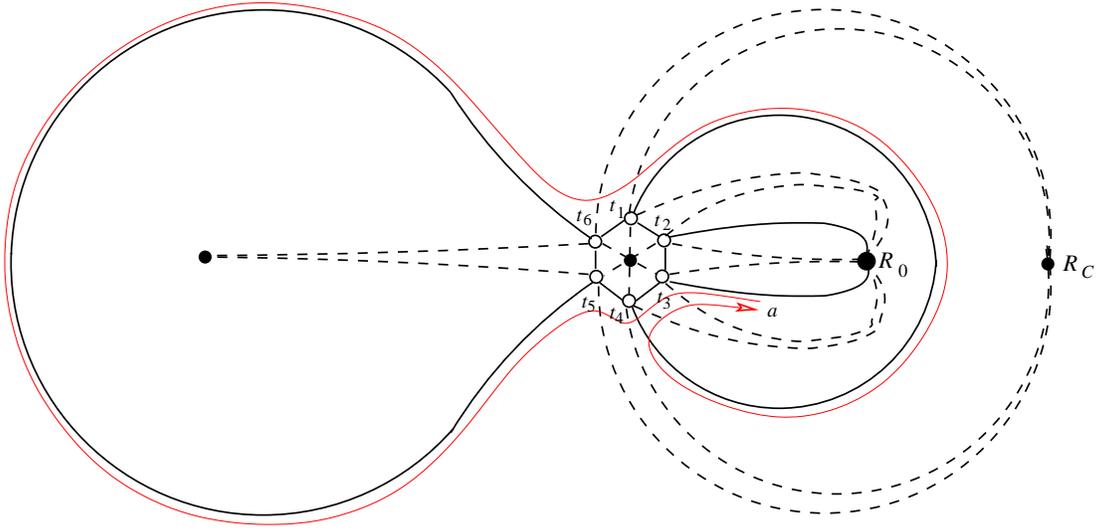}
\end{center}
\caption{Schematic presentation of the Stokes (dashed) and anti-Stokes (solid) lines in the complex plane for the RN-dS solution in four spacetime dimensions.  The open circles are the zeros and the filled circles are the poles of the function $Q^2$.  $R_C$ is the cosmological horizon and $R_0$ is the extremal horizon.  The thin arrow which starts on the anti-Stokes line that is connected to the horizon (line $a$) and ends on the same line is the path we take to determine the QNM frequency.}
\label{schem-d4-dS}
\end{figure}

\sxn{Extremal Reissner-Nordstr$\ddot{\mbox{o}}$m-de Sitter Quasinormal Modes}
\vskip 0.3cm

\begin{figure}[tb]
\begin{center}
\includegraphics[height=7cm]{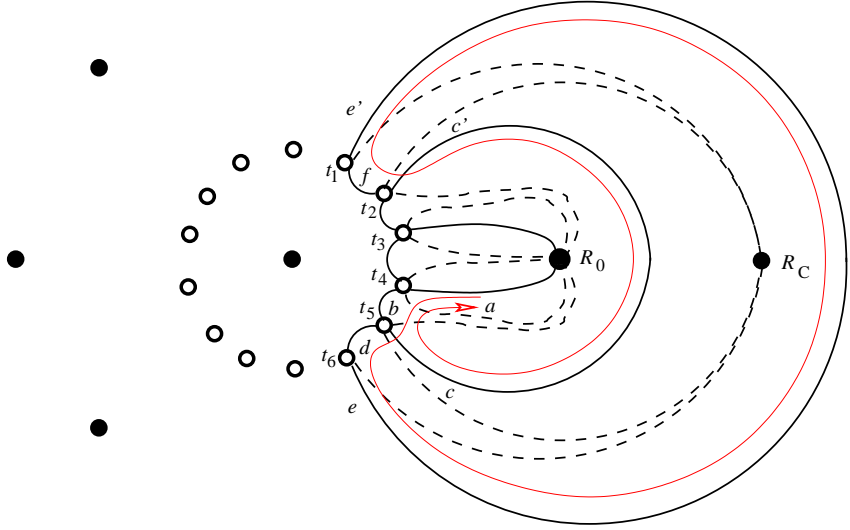}
\end{center}
\caption{Schematic presentation of the Stokes (dashed) and anti-Stokes (solid) lines in the complex plane for the RN-dS solution in six spacetime dimensions.  The open circles are the zeros and the filled circles are the poles of the function $Q^2$.  $R_C$ is the cosmological horizon and $R_0$ is the extremal horizon.  The thin arrow which starts on the anti-Stokes line that is connected to the horizon (line $a$) and ends on the same line is the path we take to determine the QNM frequency.}
\label{schem-d6-dS}
\end{figure}

In order to determine the highly damped QNM frequency of RN-dS black holes, we follow a similar methodology as explained in the previous section.  The topology of the Stokes/anti-Stokes lines for $D\ge 6$ is of the generic form shown in Fig. \ref{schem-d6-dS}, but the topology of the lines for $D=4$ and $D=5$ is different and they need to be treated differently.  In order to determine the topology, first we determine the zeros and poles of the function $Q$ in Eq. (\ref{Q^2-general}) for extremal RN-dS black holes.  In the non-extremal case the topology of these lines is shown in \cite{Natario-S}.  Using this as a guide, we can determine the topology in the extremal case. We have also confirmed our results using numerical calculations, see Fig. \ref{numerical} for the topology in $4$ and $5$ spacetime dimensions. 

For $D=4$ case, the topology of the Stokes/anti-Stokes lines is shown schematically in Fig. \ref{schem-d4-dS}.  To extract a WKB condition on the QNM frequency we start on an anti-Stokes line which is connected to the horizon (line $a$) where we should have purely in-going waves due to our boundary condition.  Therefore,
\beeq
\Psi_a = f_2^{(t_3)}.
\eneq
We follow the path shown by the thin arrow in Fig. \ref{schem-d4-dS} along the anti-Stokes lines and we return to the same place we started (line $a$).
As before, we let $\gamma_{ij}$ represent the integral of Q along the anti-Stokes lines joining consecutive zeros of $Q^2$.  These phase integrals can be evaluated using Eq. (\ref{gamma2}) where we get
\beeq
\gamma= \gamma_{12}= -\gamma_{23} = \gamma_{34} = -\gamma_{45} = -\gamma_{61} = {\pi \over 6}
\eneq
for both axial and polar perturbations.  We use a tilde (as in $\tilde{\gamma}_{14}$ and $\tilde{\gamma}_{56}$ below) to indicate that the integral is to be taken around the anti-Stokes line that circles one of the poles.  The final result is 
\begin{eqnarray}
\Psi_{a_{\rm end}} &=& \left[ -ie^{-i( \tilde{\gamma}_{14} + \tilde{\gamma}_{56})} (1+e^{2i\gamma}) \right. \nonumber \\ 
&& 
~\cdot \left.    \left(e^{-2i\gamma} + 1 + e^{2i\gamma} + e^{2i(\tilde{\gamma}_{56}-2\gamma)} + 2e^{2i(\tilde{\gamma}_{56}-\gamma)} + e^{2i\tilde{\gamma}_{56}} 
+ e^{2i(\tilde{\gamma}_{14} + \tilde{\gamma}_{56}-2\gamma)} \right)  \right] f_1^{(t_3)}
\nonumber \\ 
&-&
 \left[ e^{-i(\tilde{\gamma}_{14} + \tilde{\gamma}_{56}-2\gamma)} 
\right. \nonumber \\ 
&& 
~\cdot \left. 
\left( e^{-2i\gamma} + 1 + e^{2i\gamma} + e^{2i(\tilde{\gamma}_{56}-2\gamma)} + 2e^{2i(\tilde{\gamma}_{56}-\gamma)} + e^{2i\tilde{\gamma}_{56}} \right) \right] f_2^{(t_3)}.
\label{RN-dS-finalpsi}
\end{eqnarray}
Letting
$\Gamma_{R_0}$ be the integral of $Q$ along a simple closed curve circling the horizon at $R_0$ in a clockwise direction, and letting 
$\Gamma$ be the clockwise integral around a closed loop containing all the zeros of $Q^2$ and the poles at the origin and on the negative real axis, we then can write
\beeq
\tilde{\gamma}_{14} = \Gamma_{R_0} - \gamma_{43} - \gamma_{32} - \gamma_{21} = \Gamma_{R_0} + \gamma
\eneq
and
\beeq
\tilde{\gamma}_{56} = \Gamma -\gamma_{61}-\gamma_{12} -\gamma_{23}-\gamma_{34}-\gamma_{45} = \Gamma + \gamma ~.
\eneq
We can now rewrite Eq. (\ref{RN-dS-finalpsi}) as
\begin{eqnarray}
\Psi_{a_{\rm end}} & = & \left[ -ie^{-i(2\gamma + \Gamma + \Gamma_{R_0})} (1+e^{2i\gamma})  \right.
\nonumber \\ 
& &
~\cdot \left.  \left(e^{-2i\gamma}+ 1 + e^{2i\gamma} + e^{2i(\Gamma-\gamma)} + 2e^{2i\Gamma} + e^{2i(\gamma + \Gamma)} + e^{2i( \Gamma + \Gamma_{R_0})} \right) \right] f_1^{(t_3)}
\nonumber \\ 
&-&
 \left[ e^{-i( \Gamma + \Gamma_{R_0})} \left(e^{-2i\gamma} + 1 +  e^{2i\gamma} + e^{2i(\Gamma-\gamma)} + 2e^{2i\Gamma} + e^{2i(\gamma + \Gamma)}\right) \right] f_2^{(t_3)}~.
\label{WKB-RNdS-4d-finalpsi}
\end{eqnarray}
Applying the boundary condition that the coefficient of $f_1^{(t_3)}$ must be zero on the line $a$ which connects to $R_0$, we get the WKB condition
\beeq
[1+2\cos(2\gamma)] + [2+2\cos(2\gamma)]e^{2i\Gamma}+ e^{2i(\Gamma + \Gamma_{R_0})} = 0 ~.
\label{WKB-RNdS-4d-0}
\eneq
We now need to evaluate $\Gamma$ which contains not only the poles at the origin and on the negative real axis but also the zeros of the function $Q^2$.  It is shown in \cite{mh-4} that the solution to the wave equation at infinity is a holomorphic function for all the different types of perturbations.  This means that the monodromy of the solution at infinity is one.  In other words,
\beeq
e^{\pm i(\Gamma + \Gamma_{R_0} +\Gamma_{R_C})}=1~,
\eneq
or
\beeq
\Gamma + \Gamma_{R_0} +\Gamma_{R_C}=0~,
\label{dS-4D-Gamma-relation}
\eneq
where $\Gamma_{R_C}$ represents the integral taken in clockwise direction around the cosmological horizon at $R_C$.  Therefore, our WKB condition takes the final form
\beeq
[1+2\cos(2\gamma)] + [2+2\cos(2\gamma)]e^{-2i(\Gamma_{R_0} + \Gamma_{R_C})} +e^{-2i\Gamma_{R_C}}=0~.
\label{WKB-RNdS-4d}
\eneq
We can evaluate $\Gamma_{R_0}$ and $\Gamma_{R_C}$ using the fact that
\beeq
\Gamma_{R_0}=\oint Q dr= -2\pi i \mathop{Res}_{r=R_0} Q= -2\pi i \omega \mathop{Res}_{r=R_0} {1\over f(r)}~,
\label{Gamma-RNdS1}
\eneq
and 
\beeq
\Gamma_{R_C}=\oint Q dr= -2\pi i \mathop{Res}_{r=R_C} Q= -2\pi i \omega \mathop{Res}_{r=R_C} {1\over f(r)}~.
\label{Gamma-RNdS2}
\eneq
We cannot find $\omega$ explicitly from Eq. (\ref{WKB-RNdS-4d}) as we did in (\ref{omega_final}), instead we can check our result by taking the limit as $\lambda \rightarrow 0^+$.  In this limit, where $R_C \sim \lambda^{-1} \rightarrow +\infty$, we have
\beeq
\mathop{Res}_{r=R_C} {1\over f(r)}\sim -{1 \over {\lambda R_C}}\sim -{R_C} \rightarrow -\infty~.
\eneq
It is now easy to show that with the choice of ${\rm Re}~\omega>0$, Eq. (\ref{WKB-RNdS-4d}) reduces to
\beeq
e^{2i\Gamma_{R_0}} = -[2+2\cos(2\gamma)]= -3~,
\eneq
when $\lambda\rightarrow 0$.  This expression is exactly the same as the WKB condition we found for RN black holes in $D=4$.  Note that the choice of ${\rm Re}~\omega<0$ is not possible in this limit.  Therefore we are forced to choose a positive value for the real part of the frequency.  To get a QNM frequency with negative real part we need to take the mirror image of the path which is taken in Fig. \ref{schem-d4-dS}.

\begin{figure}[tb]
\begin{center}
\includegraphics[height=7cm]{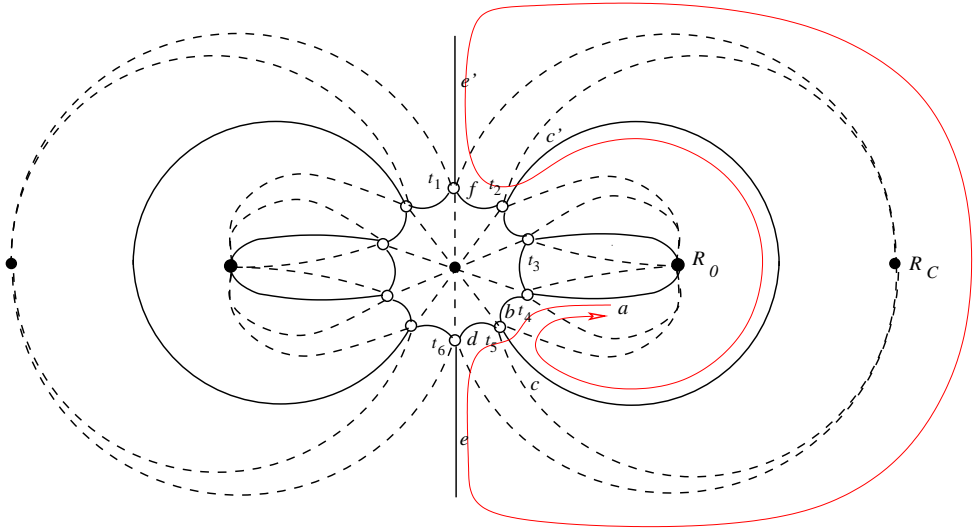}
\end{center}
\caption{Schematic presentation of the Stokes (dashed) and anti-Stokes (solid) lines in the complex plane for the RN-dS solution in five spacetime dimensions.  The open circles are the zeros and the filled circles are the poles of the function $Q^2$.  $R_C$ is the cosmological horizon and $R_0$ is the extremal horizon.  The thin arrow which starts on the anti-Stokes line that is connected to the horizon (line $a$) and ends on the same line is the path we take to determine the QNM frequency.}
\label{schem-d5-dS}
\end{figure}

In order to check further if our results are correct, we impose the WKB condition (\ref{WKB-RNdS-4d-0}) on the coefficient of $f_2^{(t_3)}$ in Eq. (\ref{WKB-RNdS-4d-finalpsi}) where we get
\begin{eqnarray}
\Psi_{a_{\rm end}} = e^{i(\Gamma+\Gamma_{R_0})}f_2^{(t_3)}=e^{-i\Gamma_{R_C}}f_2^{(t_3)}~.
\label{dS-D4-doublecheck}
\end{eqnarray}
Note that the path we take in Fig. \ref{schem-d4-dS} is equivalent to a counter-clockwise loop around the pole at $R_C$ according to the relation (\ref{dS-4D-Gamma-relation}).  Therefore, $e^{-i\Gamma_{R_C}}$ is the correct monodromy of the path taken in Fig. \ref{schem-d4-dS}.    Equation (\ref{dS-D4-doublecheck}) further confirms the validity of our results.

The topology of the Stokes/anti-Stokes lines in five dimensional spacetime is shown schematically in Fig. \ref{schem-d5-dS}.  In this topology, two of the anti-Stokes lines extend to infinity on the complex $r$-plane.  The path we take in order to extract the WKB condition on the QNM frequency is shown with a thin arrow in Fig. \ref{schem-d5-dS}.  As one can see, in order to close our loop between the zeros $t_1$ and $t_6$ we need to move along a semi-circular path to the right of the cosmological horizon $R_C$.  This path is not along any anti-Stokes line which means that the WKB solutions may change in character along this path.  To determine this change, we close this loop at a very large distance $r$ from the origin.  In this region the wave equation (\ref{Schrodinger-r}) can be approximated to be
\beeq
\frac{d^2\Psi}{dr^2}+\left[{\omega^2 \over \lambda^2 r^4}-{J_\infty^2-1 \over 4r^2}\right]\Psi=0 ~.
\label{Schrodinger-large-r}
\eneq 
The solution to this wave equation is
\beeq
\Psi(r)=A_+ \sqrt{2\pi \lambda r\over \omega} J_{J_\infty \over 2}\left({\omega\over \lambda r}\right) + A_-\sqrt{2\pi \lambda r\over \omega} J_{-{J_\infty \over 2}}\left({\omega\over \lambda r}\right)~,
\label{solutions-large-r}
\eneq 
where $J_j$ is the Bessel function of the first kind and $A_\pm$ are integration constants.  Since we are interested in the large damping limit where $\omega$ is almost purely imaginary, it is easy to see that $\omega/ \lambda r \rightarrow \infty$ on the negative imaginary axis and $\omega/ \lambda r \rightarrow -\infty$ on the positive imaginary axis.  Therefore we can use the fact that
\beeq
J_j(z)\sim \sqrt{2\over \pi z}\cos \left( z-{j\pi\over2}-{\pi\over 4}\right) \mbox{ when $z \gg 1$}~,
\label{approximate-J_j}
\eneq 
to write
\begin{eqnarray}
\Psi(r)&\sim & {\lambda r\over \omega} \left[A_+ e^{-i\alpha_+} + A_- e^{-i\alpha_-} \right]e^{i{\omega\over \lambda r}} + {\lambda r\over \omega} \left[A_+ e^{i\alpha_+} + A_- e^{i\alpha_-} \right]e^{-i{\omega\over \lambda r}}~,
\label{asymp-before}
\end{eqnarray} 
where
\begin{eqnarray}
\alpha_\pm ={\pi \over 4}(1\pm J_\infty)~.
\end{eqnarray} 
After rotating $180$ degrees from $r=\rho e^{-i\pi/2}$ (line $e$) to $r=\rho e^{i\pi/2}$ (line $e'$), where $\rho$ is a positive real number, we can write
\beeq
\sqrt{2\pi \lambda r  \over \omega} J_{\pm {J_\infty \over 2}}\left({\omega\over \lambda r }\right) \sim  2 e^{2i \alpha_\mp} \cos\left(-{\omega \over \lambda r}- \alpha_\pm \right)\mbox{ when $-{\omega \over \lambda r} \gg 1$}~.
\eneq 
Therefore, the asymptotic solution on line $e'$ can be written as
\begin{eqnarray}
\Psi(r)&\sim & -{\lambda r\over \omega} \left[A_+ e^{i(2\alpha_-+\alpha_+)} + A_- e^{i(2\alpha_+ +\alpha_-)} \right]e^{i{\omega\over \lambda r}} \nonumber \\
&& - {\lambda r\over \omega} \left[A_+ e^{i(2\alpha_- -\alpha_+)} + A_- e^{i(2\alpha_+ -\alpha_-)} \right]e^{-i{\omega\over \lambda r}}~.
\label{asymp-after}
\end{eqnarray} 
Note that, with our chosen phase of $Q$, the WKB solutions in the asymptotic region of large $r$ can be written as 
\beeq
\left\{ \begin{array}{ll}
                   f_1(r) \sim -{\lambda r\over \omega}e^{i{\omega\over \lambda r}}~,\\
                   \\
                   f_2(r)\sim -{\lambda r\over \omega}e^{-i{\omega\over \lambda r}}~.
                   \end{array}
           \right.        
\eneq
Comparing Eqs. (\ref{asymp-before}) and (\ref{asymp-after})  shows that for all types of metric perturbations ($J_\infty =4,2,0$) the coefficient of $f_1$ does not change during the rotation from line $e$ to $e'$ but the coefficient of $f_2$ reverses sign in this process.  For a similar calculation in the tortoise coordinate for $D=5$ dimensional RN-dS spacetime, the reader can refer to \cite{Natario-S}.

We now are ready to derive the WKB condition on the QNM frequency in five spacetime dimensions.  We start on an anti-Stokes line which is connected to the horizon (line $a$) where we should have purely in-going waves due to our boundary condition: 
\beeq
\Psi_a = f_2^{(t_4)}.
\eneq
We will follow the path shown in Fig. \ref{schem-d5-dS}.  In moving from $t_6$ to $t_1$, first we move along the anti-Stokes line $e$ to a large distance $r$ from the origin.  On this anti-Stokes line, the WKB solutions do not change in character.  Then we move along the big semi-circular path to the right of $R_C$ from line $e$ to $e'$.  As we explained earlier, when we take this semi-circular path the coefficient of $f_2$ reverses sign while the coefficient of $f_1$ remains unchanged.  Later, we move back to the zero $t_1$ along the line labeled $e'$.  The rest of the calculation is similar to the earlier examples.  
Using the fact that $\gamma= \gamma_{12}  = -\gamma_{23}  =\gamma_{34}  = -\gamma_{45} = \gamma_{56} = \pi/5$ in five dimensions, and 
\beeq
\tilde{\gamma}_{25} = \Gamma_{R_0} - \gamma_{54} - \gamma_{43} -\gamma_{32} = \Gamma_{R_0} - \gamma
\eneq
and
\beeq
\tilde{\gamma}_{61} = - \Gamma_{R_c} - \gamma_{12} - \tilde{\gamma}_{25} -\gamma_{56} = - \Gamma_{R_c}- \tilde{\gamma}_{25} - 2\gamma ~,
\eneq
we get the following solution when we return to line $a$:
\begin{eqnarray}
\Psi_{a_{\rm end}} & = & \left[ -ie^{-i(2\Gamma_{R_0} + \Gamma_{R_C})} (1+e^{2i\gamma})  \right.
\nonumber \\ 
& &
~\cdot \left.  \left(e^{-2i\gamma}+ 2 + e^{2i\gamma} + e^{2i\Gamma_{R_0}} - e^{2i(\Gamma_{R_0}+\Gamma_{R_C}-\gamma)} - e^{2i(\Gamma_{R_0}+\Gamma_{R_C})} - e^{2i(\gamma + \Gamma_{R_0}+\Gamma_{R_C})}  \right) \right] f_1^{(t_4)}
\nonumber \\ 
&-&
e^{-i( 2\Gamma_{R_0} + \Gamma_{R_C})} \left(e^{-2i\gamma}+ 2 + e^{2i\gamma} 
\right.
\nonumber \\ 
& & ~~~~~~~~~~~~~~~~~~\left. 
- e^{2i(\Gamma_{R_0}+\Gamma_{R_C}-\gamma)}- e^{2i(\Gamma_{R_0}+\Gamma_{R_C})} - e^{2i(\gamma + \Gamma_{R_0}+\Gamma_{R_C})}  \right) f_2^{(t_4)}~.
\label{dS5}
\end{eqnarray}
Here $\tilde{\gamma}_{25}$ and $\tilde{\gamma}_{61}$ are the phase integrals along the path taken to the right of the poles at $R_0$ and $R_C$ respectively.
Due to the boundary condition at the horizon, we clearly must require that the coefficient of $f_1^{(t_4)}$ in $\Psi_{a_{\rm end}}$ to be zero.  This gives the WKB condition
\beeq
[2+2\cos(2\gamma)] + e^{2i\Gamma_{R_0}}- [1+2\cos(2\gamma)]e^{2i(\Gamma_{R_0} + \Gamma_{R_C})} =0~.
\label{WKB-RNdS-5d}
\eneq
As in the four spacetime dimension, it is easy to show that when $\lambda \rightarrow 0$ the above condition reduces to the RN WKB condition (\ref{RN-WKB}).
In order to further confirm the validity of our results, we impose the WKB condition (\ref{WKB-RNdS-5d}) on the coefficient of $f_2^{(t_4)}$ in (\ref{dS5}) where we get
\begin{eqnarray}
\Psi_{a_{\rm end}} =e^{-i\Gamma_{R_C}}f_2^{(t_3)}~.
\end{eqnarray}
$e^{-i\Gamma_{R_C}}$ is the correct monodromy of the path we take in the counter-clockwise direction around $R_C$ as shown in Fig. \ref{schem-d5-dS}.

Finally, in Fig. \ref{schem-d6-dS}, we show the schematic behavior of the Stokes/anti-Stokes lines in six spacetime dimensions, which turns out to be generic for every dimension greater than five\cite{Natario-S}.  Therefore the following results are valid for spacetime dimensions $D\ge 6$.

Again we start on line $a$ where 
\beeq
\Psi_a = f_2^{(t_4)}.
\label{dS1}
\eneq
We follow the path shown in Fig. \ref{schem-d6-dS} by the thin arrow and we return to line $a$.  The final result is
\begin{eqnarray}
\Psi_{a_{\rm end}} & = & \left[ -ie^{-i(2\Gamma_{R_0} + \Gamma_{R_C})} (1+e^{2i\gamma})  \right.
\nonumber \\ 
& &
~\cdot \left.  \left(e^{-2i\gamma}+ 2 + e^{2i\gamma} + e^{2i\Gamma_{R_0}} + e^{2i(\Gamma_{R_0}+\Gamma_{R_C}-\gamma)} + e^{2i(\Gamma_{R_0}+\Gamma_{R_C})} + e^{2i(\gamma + \Gamma_{R_0}+\Gamma_{R_C})}  \right) \right] f_1^{(t_4)}
\nonumber \\ 
&-&
 \left[ e^{-i( 2\Gamma_{R_0} + \Gamma_{R_C})} \right.
\nonumber \\ 
& &
~\cdot \left. 
 \left(e^{-2i\gamma}+ 2 + e^{2i\gamma} + e^{2i(\Gamma_{R_0}+\Gamma_{R_C}-\gamma)} + e^{2i(\Gamma_{R_0}+\Gamma_{R_C})} + e^{2i(\gamma + \Gamma_{R_0}+\Gamma_{R_C})}  \right) \right] f_2^{(t_4)}~.
\label{dS2}
\end{eqnarray}
Comparing Eqs. (\ref{dS1}) and (\ref{dS2}), we must require the coefficient of $f_1^{(t_4)}$ in $\Psi_{a_{\rm end}}$ to be zero, which gives the WKB condition
\beeq
[2+2\cos(2\gamma)] + e^{2i\Gamma_{R_0}}+ [1+2\cos(2\gamma)]e^{2i(\Gamma_{R_0} + \Gamma_{R_C})} =0~,
\label{WKB-RNdS-6d}
\eneq
where $\gamma={(n-1)\over 2(2n-1)}\pi$ for all different types of metric perturbations as was shown in the previous section.  Note that the WKB condition (\ref{WKB-RNdS-6d}) is identical to the condition (\ref{WKB-RNdS-4d}) that we found for four spacetime dimensions.  Therefore, we conclude that the condition (\ref{WKB-RNdS-6d}) on the QNM frequency is valid for $D=4$ and $D\ge6$ spacetime dimensions.  In the limit $\lambda \rightarrow 0$, the above WKB condition reduces to the RN condition (\ref{RN-WKB}).  Once again, in order to further confirm the validity of our results, we can impose the WKB condition (\ref{WKB-RNdS-4d}) on the coefficient of $f_2^{(t_4)}$ in (\ref{dS2}) where we get
\begin{eqnarray}
\Psi_{a_{\rm end}} =e^{-i\Gamma_{R_C}}f_2^{(t_3)}~.
\end{eqnarray}
$e^{-i\Gamma_{R_C}}$ is the correct monodromy of the counter-clockwise path we take around the cosmological horizon $R_C$ in Fig. \ref{schem-d6-dS}. 

The results of this section match exactly with the extremal limit of the results found by Natario and Schiappa\cite{Natario-S} for the highly damped QNMs of non-extremal RN-dS black holes.

\sxn{ Conclusions}
\vskip 0.3cm

We have calculated explicitly the QNM frequencies of extremal RN and RN-dS black holes in arbitrary dimensions ($D\ge 4$) using the analytic technique of Andersson and Howls \cite{Andersson}.  We have shown that the highly damped QNMs of extremal RN and RN-dS black holes match exactly with the extremal limit of the non-extremal black hole QNMs previously obtained in \cite{Andersson} and \cite{Natario-S}.  Our analytical results for extremal RN black holes agrees very well with the numerical results obtained by Berti in \cite{Berti1}.

Is it possible to link the results of this paper to quantum gravity via Hod's conjecture\cite{Hod}?  Unfortunately we do not have a good answer to this question.  At first, the appearance of the famous $\ln(3)$ in the RN QNM frequency in four dimensional spacetime seems promising, but then it turns out that in higher spacetime dimensions we do not get $\ln(3)$.  Instead, we get the natural logarithm of a real number between $2$ and $3$.  Of course, as it is argued in \cite{Andersson}, this is not very surprising since the general perturbation of a RN black hole corresponds to a mixture of electromagnetic and gravitational waves.  Contrary to gravitational waves which are the oscillations of spacetime itself, electromagnetic waves propagate in a fixed background.  This mixture may prevent a simple correspondence between the highly damped QNMs of black holes and their quantum area spectrum.  If we assume that such a simple correspondence does exist\cite{Andersson,Setare}, one may raise the question that why we do not have the natural logarithm of an integer in higher dimensional spacetimes.  One answer to this question could be that, in extremal black holes, entropy is argued\cite{Hawking-Horowitz} to be zero which means that we do not need to impose the statistical interpretation\cite{Hod, Bekenstein-1} of the Bekenstein-Hawking entropy spectrum where we need to have the natural logarithm of an integer.  If it turns out that extremal black holes have a non-zero entropy then this argument will lose its validity.  In the case of RN-dS black holes, the WKB conditions on QNM frequencies become more complicated due to the presence of a cosmological constant.  It is again not surprising to see that the presence of a cosmological constant modifies the black hole QNMs further which could potentially prevent a simple correspondence between the highly damped QNMs of black holes and their quantum area spectrum.  In the limit $\lambda \rightarrow 0$, we showed that the results for extremal RN-dS black hole QNMs reduces to the results we obtained for the RN case.

One more question which can be raised is related to the fact that extremal black hole QNMs are the limiting case of the non-extremal black hole QNMs.  Is this result provide an evidence against the arguments of Hawking {\it et al} \cite{Hawking-Horowitz} that extremal black holes are not a limiting case of non-extremal black holes since the entropy changes discontinuously when the extremal limit is reached, or is this result an indication that QNMs do not have a lot to do with black hole thermodynamics and quantum mechanics.  We do not have answers to these questions.

A puzzling issue in the results of this paper is related to the fact that the WKB condition for QNMs in $D=5$ dimensional RN-dS spacetime is different than the other dimensions.  This raises the question of what is special about $D=5$.  This issue needs to be investigated further.
  
For the future, we are planing to calculate the highly damped QNMs of extremal Reissner-Nordstr$\ddot{\rm o}$m-anti de Sitter (RN-AdS) black holes.  These black holes are especially interesting in the context of AdS/CFT correspondence\cite{Maldacena}.  Considering the results of this paper, we strongly believe that the highly damped QNM frequencies of extremal RN-AdS black holes would match exactly with the extremal limit of the non-extremal RN-AdS black hole QNM frequencies.  This issue is under investigation.

\vskip .5cm

\leftline{\bf Acknowledgments}
We would like to thank Samir Mathur for motivating us in initiating this project.  We also would like to thank Axel Boldt for the invaluable discussions we had at the beginning of the project and Gabor Kunstatter for his suggestions on our concluding remarks.


\def\jnl#1#2#3#4{{#1}{\bf #2} (#4) #3}

\def\Zphys{{\em Z.\ Phys.} }
\def\jssc{{\em J.\ Solid State Chem.\ }}
\def\jpsJ{{\em J.\ Phys.\ Soc.\ Japan }}
\def\ptps{{\em Prog.\ Theoret.\ Phys.\ Suppl.\ }}
\def\PTP{{\em Prog.\ Theoret.\ Phys.\  }}
\def\LNC{{\em Lett.\ Nuovo.\ Cim.\  }}

\def\JMP{{\em J. Math.\ Phys.} }
\def\NPB{{\em Nucl.\ Phys.} B}
\def\NP{{\em Nucl.\ Phys.} }
\def\PLB{{\em Phys.\ Lett.} B}
\def\PL{{\em Phys.\ Lett.} }
\def\PRL{\em Phys.\ Rev.\ Lett. }
\def\PRB{{\em Phys.\ Rev.} B}
\def\PRD{{\em Phys.\ Rev.} D}
\def\PR{{\em Phys.\ Rev.} }
\def\PRe{{\em Phys.\ Rep.} }
\def\AP{{\em Ann.\ Phys.\ (N.Y.)} }
\def\RMP{{\em Rev.\ Mod.\ Phys.} }
\def\ZPC{{\em Z.\ Phys.} C}
\def\SCI{\em Science}
\def\CMP{\em Comm.\ Math.\ Phys. }
\def\MPLA{{\em Mod.\ Phys.\ Lett.} A}
\def\IJMPB{{\em Int.\ J.\ Mod.\ Phys.} B}
\def\cmp{{\em Com.\ Math.\ Phys.}}
\def\JPA{{\em J.\  Phys.} A}
\def\CQG{\em Class.\ Quant.\ Grav.~}
\def\ATMP{\em Adv.\ Theoret.\ Math.\ Phys.~}
\def\PRSA{{\em Proc.\ Roy.\ Soc.\ Lond.} A }
\def\IJTP{\em Int.\ J.\ Theor.\ Phys.~}
\def\ibid{{\em ibid.} }
\vskip 1cm

\leftline{\bf References}

\renewenvironment{thebibliography}[1]
        {\begin{list}{[$\,$\arabic{enumi}$\,$]}  
        {\usecounter{enumi}\setlength{\parsep}{0pt}
         \setlength{\itemsep}{0pt}  \renewcommand{\baselinestretch}{1.2}
         \settowidth
        {\labelwidth}{#1 ~ ~}\sloppy}}{\end{list}}


\end{document}